\input {epsf}

\documentclass [12pt] {article}

\usepackage [cp1251] {inputenc}

\usepackage {graphicx}
\input {epsf}

\begin {document}
\author {Kozlov G.G.}
\title {To the possibility of a mobility edge in a random 1D lattice with long-range correlated disorder}
\maketitle
\begin {abstract}
Tight-binding 1D random system with long-range correlations is studied numerically 
using the localisation criterium, which represents the number of sites, covered by 
the wave function. At low degrees of disorder the signs of a mobility edge, predicted 
in \cite{Izr}, were found. The possibility of exact mobility edge in the system under 
consideration is discussed. 
\end {abstract}

\section * {Introduction}

 Absence of a mobility edge in 1D disordered systems with short-range correlations is well 
established \cite{Lif}. Despite the fact that this statement is proved  only for short-range 
correlated systems,  the occurrence  of a   mobility edge in 1D random system with some particular 
long-range correlated disorder, reported in \cite{Izr}, seems to be very interesting and unexpected.  
 The authors of \cite{Izr} have managed to construct the correlated random potential for which the inverse localisation length (calculated theoretically for small degrees of disorder) was found to be zero for the central half-band of the energy spectrum and to be a non-zero constant for the remaining half-band.
From the author's of this note point of view,  the infinitiness of localisation length is not an 
unambiguous criterium of localisation. For example, the zero-state in 1D chain with off-diagonal 
disorder has infinite localisation length \cite{The} but, nevertheless, zero-state 
was found to be localised in  sense of the  basic Anderson criterium  of localisation \cite{Koz2}. Therefore  the       verification of the result obtained in \cite{Izr} by means of Anderson (or equivalent)  criterium of localisation is necessary. 
The main task of this note is to put forward the additional arguments in 
favor of presence of a mobility edge in  correlated 1D random chain,  
constructed by recipe \cite{Izr}. To do this we  perform the numerical  exploration of   
eigen vectors of the random matrix of an appropriate   disordered Hamiltonian by means of  
a criterium of localisation, differing from the {\it finiteness of the localisation length}.

\section{ 1D long-range correlated random chain: the Hamiltonian.}

The matrix of the Hamiltonian  we are going to study has the tight-binding form
\begin{equation}
H_{rr'}=\delta_{rr'}\varepsilon_r+\delta_{r,r'+1}+\delta_{r,r'-1},\hskip10mm r,r,=1,2,...,N
\label{0}
\end{equation}
with {\it long-range correlated} random potential $\varepsilon_r$, generated in accordance 
with the following algorithm \cite{Izr}:
\begin{equation}
\varepsilon_r=\sum_{m=-\infty}^\infty G(r-m)\xi_m \hskip15mm
G(m)=\sqrt{2\over 3}{3\over 2\pi m}\sin\bigg [{2\pi m\over 3}\bigg]
\label{00}
\end{equation}
w	here $\xi_m$ are similarly destributed independent random numbers with $\langle \xi_m\rangle=0$
 and $\langle \xi_m^2\rangle=d^2/2$ for any integer $m\in [-\infty,+\infty]$. The 
degree of disorder is controlled by $d$. As it was shown in \cite{Izr}, the spectral 
dependence of the inverse localisation length calculated for the Hamiltonian   (\ref{0})  
with random potential  (\ref{00}) at small degree of disorder $d$, goes down to zero 
in the central half-band $E\in [-1,1]$  and remains constant at $E\in [\pm 2,\pm 1]$. 
The authors of \cite{Izr} have interpreted this as the occurrence of  a  mobility edges at $E=\pm 1$. 
Below we verify this statement from the view point of our criterium of localisation, which is 
equivalent to the Anderson's criterium.

\section{Number of sites covered by the wave function: the criterium of localisation}

We will start from reminding of a simple criterium of localisation, suggested in \cite{Koz2}, which we will use below.
 We will characterise an arbitrary state $\bf \Psi$ by a  {\it number $N^\ast$ of sites, covered by an appropriate wave function $\bf\Psi$,} calculated as  follows.
 
  As $\bf\Psi$ is the eigen vector of  a random Hamiltonian matrix in site representation, $\bf\Psi$ 
represents  a vector-cloumn with componets $\Psi_1,..,\Psi_i,..,\Psi_N$ where $i$ is the site index.
   If $|\Psi_i|^2 = 0$ at a certain site $i$, then this site is {\it not covered} by a state $\bf \Psi$ and is
   not taken into account  while calculating $N^\ast$. Vice versa, if $|\Psi_i|^2$ reaches its maximum
  $|\Psi|^2_{max}$ at the site $i$, then this site is {\it covered completely} and its contribution while calculating
  $N^\ast $ is equal to unit. In the general case, the contribution of an arbitrary site $i$ is equal to
  $|\Psi_i|^2/|\Psi |^2_{max} $ and we obtain the following formula for  $ N^\ast $:
   \begin{equation}
  N^\ast =N^\ast\{{\bf\Psi\}}=\sum_i^N {|\Psi_i|^2\over |\Psi|^2_{max}}={1
  \over |\Psi|^2_{max}}
  \end{equation}
 We can now introduce energy  depending $N^\ast(E)$   as follows. 
 Let us specify some energy interval $dE<<E_{max}-E_{min}$ where $E_{max}-E_{min}$  is  a typical range of eigen energies for the random Hamiltonian under consideration. 
 Consider all states with energies within the interval $[E,E+dE]$ and denote by $\sum_{E,E+dE}$  the summation over all these states. Now calculate $N^\ast(E)$ as the following average value over all states with energies within the interval $[E,E+dE]$:
 \begin{equation}
 \langle N^\ast (E)\rangle=\bigg\langle \sum_{E,E+dE} N^\ast ({\bf\Psi})/\sum_{E,E+dE} 1\bigg\rangle
 \label{2}
 \end{equation}
 with $\langle \rangle$ standing for the averaging over realisations of disorder. 
For small enough $dE$ the function (\ref{2}) does not depend on $dE$.
  For the Hamiltonian represented by a random matrix of size $N$, the 
function $\langle N^\ast (E)\rangle $ in the spectral range of localised states 
should not depend on $N$ (if $N$  is large enough for $\langle N^\ast (E)\rangle <N$  in this spectral
   region). In the spectral range of delocalised states (if it is exist), the function $\langle N^\ast (E)\rangle $ must increase  as $\sim N$. This property of $\langle N^\ast (E)\rangle$ one can use as a criterium for  selection of localised and delocalised states in numerical experiments.  
 
 \section{Results }
 
 \begin{figure}
  \includegraphics[width=.8\columnwidth,clip]{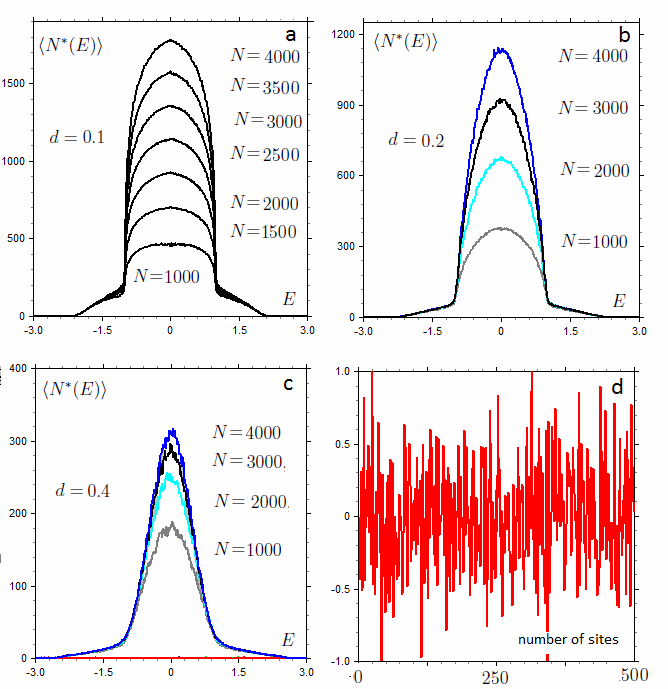}
  \caption{ Panels (a),(b) and (c) -- the spectral dependences of $\langle N^\ast(E)\rangle$  for 
various sizes $N=1000,..., 4000$  of a random Hamiltonian matrix (\ref{0}) 
at variuos degrees of disorder $d=0.1, 0.2, 0.4$. The matrix size $N$ and the 
degree of disorder $d$ are shown at the panels. In all cases the averaging over 
100 realisations was performed, $dE=0.012$. Panel (d) shows the realisation of a random potential (\ref{00}).}
  \label{fig1}
  \end{figure}

 In our calculations the auxiliary random values $\xi_m$ were of gauss type with 
the distribution function in the form 
 \begin{equation}
 \rho(\xi)={1\over\sqrt\pi d}e^{-(\xi/d)^2}
 \end{equation}
  The spectral dependences of $\langle N^\ast(E)\rangle$  for 
various sizes $N=1000,..., 4000$ of a random Hamiltonian matrix (\ref{0}) at variuos degrees 
of disorder $d=0.1, 0.2, 0.4$ are presented at Fig.\ref{fig1}. From Fig.\ref{fig1}a  one can see that, 
at small disorder $d=0.1$, the well-pronounced signs of delocalised states 
in the central half-band $E\in [-1,1]$ are observed. 
In this spectral range the number $\langle N^\ast(E)\rangle$ of sites covered by the 
wave functions is proportional to the size $N$ of a random matrix (\ref{0}). 
At $E\in [\pm 1,\pm 2]$ the number $\langle N^\ast (E)\rangle$ does not 
depend on $N$. As it was mentioned above, such a behaviour corresponds to the localised character of states 
at $E\in [\pm 1,\pm 2]$.
 So, the signs of  a mobility edge at $E=\pm 1$, predicted in \cite{Izr}, are observed.
 
Despite the fact that the above data have confirmed the statements of \cite{Izr} there are at least 
two questions remain to be answered.
The first question can be formulated as follows. In accordance with \cite{Izr} the localisation 
length $L_{loc}$ of  states with   energies in the  range $E\in [\pm 1,\pm 2]$ {\it must be constant}. 
 But our calculations  
show that the number of sites $\langle N^\ast (E)\rangle$ 
 covered by the localised wave functions is strongly  depend on 
energy in this energy interval (Fig.\ref{fig1}a), varying from 7 -- 8 (nearly zero) at $E=\pm 2$ to $\sim 200$ 
at the "mobility edge'' $E=\pm 1$.
 It seems strange because both these quantities ($\langle N^\ast (E)\rangle$ and $L_{loc}$) must describe the 
size of localised states and their energy dependence  must be similar $L_{loc}\sim \langle N^\ast (E)\rangle $.  
 In our opinion this inconsistency   reveals the ambiguous and disputable  
character of the localisation length  as a measure of localisation of states.

 The second question  relates to the behaviour of the above random model  
at large degrees of disorder. The existence of a mobility edge (in the sense of inverse 
localisation length)  was obtained  in \cite{Izr} in the limit of small disorder $d\rightarrow 0$.  
Whether it is possible to extend this result for the case of an 
arbitrary strong disorder?
 To clear up this we have studied the behaviour of $\langle N^\ast (E)\rangle$ at large  
degrees of disorder. The results are presented at Fig.\ref{fig1} (b,c). 
 One can see that Fig.\ref{fig1} (b,c) give grounds to conclude that at large 
disorder {\it all states of the system are localised}:
  the number of covered sites $\langle N^\ast (E)\rangle$ for the  states at the center 
of the band saturate, i.e. the nearly linear  dependence $\langle N^\ast (E)\rangle$  
on $N$, which takes place for $d=0.1$ (Fig.\ref{fig1}a),   violates.
 Having this in mind, one can suppose that  if we had an opportunity to make  
calculations for $d=0.1$ (Fig.\ref{fig1}a) with matrixes of the 
 size greater than 4000, we would also observe the  saturation of linear 
dependence of $\langle N^\ast (E)\rangle$  on $N$.
  Or  there is some critical degree of disorder $d_c$, such that for $d>d_c$, 
 the states in the center of the band become localised? 

In our opinion all these questions are still open.

Even if the further exploration of the correlated random system (\ref{0}), (\ref{00}) will 
disprove the existence of exact mobility edge in this system, the 
curious spectral dependence of a  number of covered sites $\langle N^\ast (E)\rangle$ with a 
"mobility edge'' (may be, virtual), looks very interesting.

\end {document}